\documentstyle[amssymb,preprint,aps]{revtex}

\begin{document}
\draft
\title{Soliton Gauge States and T-duality of Closed Bosonic String Compatified on
Torus}
\author{Jen-Chi Lee\thanks{
e-mail: jclee@cc.nctu.edu.tw; On leave of absence from : Department of
Electrophysics, National Chiao Tung University, Taiwan. }}
\address{Department of physics, MIT, 77 Massachusetts Avenue, Cambridge, MA 02139, USA}
\maketitle

\begin{abstract}
We study soliton gauge states in the spectrum of bosonic string compatified
on torus. The enhanced Kac-Moody gauge symmetry, and thus T-duality, is
shown to be related to the existence of these soliton gauge states in some
moduli points.
\end{abstract}

\pacs{}

\pagestyle{plain}


\widetext

\section{Introduction}

\label{sec:intro}String duality \cite{1} has been the subject of active
research for the last few years. The five consistent perturbative string
theories are now known to be related to each other through various duality
symmetries. It is believed that they are merely different moduli points of a
single underlying theory termed M-theory. The best known string duality is
the T-duality which can be understood perturbatively \cite{2}. T-duality
relates a string theory in a background with large volume to another string
theory in a background with small volume. For example, it has been shown
that the Heterotic $E_{8}\otimes E_{8}$ and $SO(32)$ theories sit at
different points. which are T-dual to each other, of the moduli space of the
same Heterotic theory below ten dimension \cite{3}.

For the compatified bosonic string, the discrete T-duality group were shown
to be the residual Weyl subgroup of the enhanced Kac-Moody gauge symmetry 
\cite{2}. On the other hand, it has been known that space-time gauge
symmetry of uncompatified string is related to the existence of gauge states
in the spectrum \cite{4}. For the 10D Heterotic string, the Heterotic gauge
states \cite{5} are responsible for the massless $E_{8}\otimes E_{8}$ or $%
SO(32)$ gauge symmetry and are used to predict the existence of an infinite
number of massive Einstein-Yang-Mills type gauge symmetry. For the toy 2D
string, the discrete gauge states \cite{6} are responsible for the $%
w_{\infty }$ symmetry of the Liouville theory. It is thus of interest to
understand the gauge state structure of the compatified string theory, and
study their relation to the enhanced Kac-Moody gauge symmetry.

In this paper, for simplicity, we will study gauge states of closed bosonic
string compatified on torus. In addition to the usual gauge states, we will
discover soliton gauge states (SGS) in the spectrum of some moduli points.
These gauge states and SGS form a realization of enhanced Kac-Moody gauge
symmetry group in the gauge state sector of the spectrum. Since T-duality
group is the Weyl subgroup of the enhanced gauge group, SGS can be
considered as the origin of the discrete T-duality group. In section II, we
derive massless gauge states of bosonic string compatified on $R^{25}\otimes
T^{1}$ at self-dual point $R=\sqrt{2}$, and show that they form a
representation of the enhanced $SU(2)_{R}\otimes SU(2)_{L}$ gauge group. In
section III, we generalize the calculation to $R^{26-D}\otimes T^{D}$ and
give examples at some moduli points. Section IV is devoted to the discussion
of massive SGS. We will find that there is an infinite number of massive SGS
which exists at some moduli points. The existence of these massive SGS
implies that there is an infinite enhanced gauge symmetry of compatified
string theory. Finally a brief conclusion is given in section V.

\section{Soliton Gauge State on $R^{25}\otimes T^{1}$}

In the simplest torus compatification, one coordinate of the string was
compatified on a circle of radius $R$ \cite{7}

\begin{equation}
X^{25}\left( \sigma +2\pi ,\pi \right) =X^{25}\left( \sigma ,\pi \right)
+2\pi R_{n}  \eqnum{2.1}
\end{equation}
Singlevaluedness of the wave function then restricts the allowed momenta to
be $p^{25}=m/R$ with $m,n\in Z$. The mode expansion of the compatified
coordinate for right (left) mover is

\begin{eqnarray}
X_{R}^{25} &=&\frac{1}{2}x^{25}+\left( p^{25}-\frac{1}{2}nR\right) \left(
\tau -\sigma \right) +i\sum_{r\neq 0}\frac{1}{r}\alpha _{r}^{25}e^{-ir\left(
\tau -\sigma \right) }  \eqnum{2.2} \\
X_{L}^{25} &=&\frac{1}{2}x^{25}+\left( p^{25}+\frac{1}{2}nR\right) \left(
\tau +\sigma \right) +i\sum_{r\neq 0}\frac{1}{r}\stackrel{\thicksim }{\alpha 
}_{r}^{25}e^{-ir\left( \tau +\sigma \right) }  \eqnum{2.3}
\end{eqnarray}
We have normalized the string tension to be $\frac{1}{4\pi T}=1$ or $\alpha
^{\prime }=2$ .\mathstrut The Virasoro operators can be written as

\begin{eqnarray}
L_{0} &=&\frac{1}{2}\left( p^{25}-\frac{1}{2}nR\right) +\frac{1}{2}p^{\mu
^{2}}+\sum_{n=1}^{\infty }\alpha _{-n}\cdot \alpha _{n}  \eqnum{2.4} \\
\stackrel{\thicksim }{L_{0}} &=&\frac{1}{2}\left( p^{25}+\frac{1}{2}%
nR\right) +\frac{1}{2}p^{\mu ^{2}}+\sum_{n=1}^{\infty }\stackrel{\thicksim }{%
\alpha }_{-n}\cdot \stackrel{\thicksim }{\alpha }_{n}  \eqnum{2.5}
\end{eqnarray}
and

\begin{eqnarray}
L_{m} &=&\frac{1}{2}\alpha _{0}^{2}+\sum_{-\infty }^{\infty }\alpha
_{m-n}\cdot \alpha _{n}  \eqnum{2.6} \\
\stackrel{\thicksim }{L}_{m} &=&\frac{1}{2}\stackrel{\thicksim }{\alpha }%
_{0}^{2}+\sum_{-\infty }^{\infty }\stackrel{\thicksim }{\alpha }_{m-n}\cdot 
\stackrel{\thicksim }{\alpha }_{n}\text{ }\left( m\neq 0\right)  \eqnum{2.7}
\end{eqnarray}
where

\begin{eqnarray}
\alpha _{0}^{25} &=&p^{25}-\frac{1}{2}nR\equiv p_{R}^{25}  \eqnum{2.8} \\
\stackrel{\thicksim }{\alpha }_{0}^{25} &=&p^{25}+\frac{1}{2}nR\equiv
p_{L}^{25}  \eqnum{2.9}
\end{eqnarray}
and the 25d momentum is $\alpha _{0}^{\mu }=\stackrel{\thicksim }{\alpha }%
_{0}^{\mu }=p^{\mu }\equiv k^{\mu }$. In the old covariant quantization of
the theory, in addition to the physical propagating states, there are four
types of gauge states in the spectrum

\begin{equation}
I.a\left| \psi \right\rangle =L_{-1}\left| \chi \right\rangle \text{ where }%
L_{m}\left| \chi \right\rangle =0,\left( \stackrel{\thicksim }{L}_{m}-\delta
_{m}\right) \left| \chi \right\rangle =0,\text{ }\left( m=0,1,2,\ldots
\right)  \eqnum{2.10}
\end{equation}

\begin{equation}
II.a\left| \psi \right\rangle =\left( L_{-2}+\frac{3}{2}L_{-1}^{2}\right)
\left| \chi \right\rangle \text{ where }\left( L_{m}+\delta _{m}\right)
\left| \chi \right\rangle =0,\left( \stackrel{\thicksim }{L}_{m}-\delta
_{m}\right) \left| \chi \right\rangle =0,\text{ }\left( m=0,1,2,\ldots
\right)  \eqnum{2.11}
\end{equation}
and by interchanging all left and right mover operators, one gets $I.b$ and $%
II.b$ states. Type II states are zero-norm gauge states only at critical
space-time dimension. We will only calculate type a states. Similar results
can be easily obtained for type $b$ states. For type $I.a$ state, the $m=0$
constraint of Eq. (2.10) gives

\begin{equation}
M^{2}=\frac{m^{2}}{R^{2}}+\frac{1}{4}n^{2}R^{2}+N+\stackrel{\thicksim }{N}-1
\eqnum{2.12}
\end{equation}

\begin{equation}
N-\stackrel{\thicksim }{N}=mn-1  \eqnum{2.13}
\end{equation}
where $N\equiv \sum\limits_{n=1}^{\infty }\alpha _{-n}\cdot \alpha _{n}$ and 
$\stackrel{\thicksim }{N\text{ }}\equiv \sum\limits_{n=1}^{\infty }\stackrel{%
\thicksim }{\alpha }_{-n}\cdot \stackrel{\thicksim }{\alpha }_{n}$. For
massless $M^{2}=0$ states, $N+\stackrel{\thicksim }{N}=0$ or $1$. The
solutions of Eqs. (2.12) and (2.13) are

\begin{equation}
N=0,\stackrel{\thicksim }{N}=1,m=n=0\text{ ({\it any R})}  \eqnum{2.14}
\end{equation}
or

\begin{equation}
N=\stackrel{\thicksim }{N}=0,m=n=\pm 1,R=\sqrt{2}  \eqnum{2.15}
\end{equation}
Equation (2.15) gives us our first SGS. It is easy to write down the
explicit form of $\left| \chi \right\rangle $ and $\left| \psi \right\rangle 
$, and impose the $m\neq 0$ constraints of Eq. (2.10). There are also a
vector and a scalar gauge states in Eq. (2.14). Similar results can be
obtained for the type $I.b$ state. In this case, $m=-n=\pm 1$. There is no
type II solution in the massless case. We note that there are massless
soliton gauge states only when $R=\sqrt{2}$ which is known as self-dual
point in the moduli space. The vertex operators of all gauge states are
calculated to be

\begin{equation}
k_{\mu }\theta _{\nu }\partial X_{R}^{\mu }\stackrel{\_}{\partial }%
X_{L}^{\nu }e^{ikx};\text{ }L\leftrightarrow R  \eqnum{2.16}
\end{equation}

\begin{eqnarray}
&&k_{\mu }\partial X_{R}^{\mu }\stackrel{\_}{\partial }X_{L}^{25}e^{ikx} 
\eqnum{2.17} \\
&&k_{\mu }\stackrel{\_}{\partial }X_{L}^{\mu }\partial X_{R}^{25}e^{ikx} 
\eqnum{2.18} \\
&&k_{\mu }\partial X_{R}^{\mu }e^{\pm i\sqrt{2}X_{L}^{25}}e^{ikx} 
\eqnum{2.19} \\
&&k_{\mu }\partial X_{L}^{\mu }e^{\pm i\sqrt{2}X_{R}^{25}}e^{ikx} 
\eqnum{2.20}
\end{eqnarray}
It is easy to see that the three gauge states of Eqs. (2.18) and (2.20) form
a representation of $SU(2)_{R}$ Kac-Moody algebra. Similarly, Eqs. (2.17)
and (2.19) form a repersentation of $SU(2)_{L}$ Kac-Moody algebra. The
vector gauge states in Eq. (2.16) are responsible for the gauge symmetry of
graviton and antisymmetric tensor field. We see that the self-dual point $R=%
\sqrt{2}$ is very special even from the gauge sector point of view.

\section{Soliton Gauge State on $R^{26-D}\otimes T^{D}$}

In this section we compatify D coordinates on a D-dimensional torus $%
T^{D}\equiv \frac{R^{D}}{2\pi \Lambda ^{D}}$

\begin{equation}
\stackrel{\rightarrow }{X}\left( \sigma +2\pi ,\pi \right) =\stackrel{%
\rightarrow }{X}\left( \sigma ,\pi \right) +2\pi \stackrel{\rightarrow }{L} 
\eqnum{3.1}
\end{equation}
with

\mathstrut 
\begin{equation}
\stackrel{\rightarrow }{L}=\sum_{i=1}^{D}n_{i}\left( R_{i}\frac{\stackrel{%
\rightarrow }{e}_{i}}{\sqrt{2}}\right) \in \left( \Lambda ^{D}\right) 
\eqnum{3.2}
\end{equation}
where $\Lambda ^{D}$ is a D-dimensional lattice with a basis $\left\{ R_{1}%
\frac{\stackrel{\rightarrow }{e}_{1}}{\sqrt{2}},R_{2}\frac{\stackrel{%
\rightarrow }{e}_{2}}{\sqrt{2}},\ldots ,R_{D}\frac{\stackrel{\rightarrow }{e}%
_{D}}{\sqrt{2}}\right\} $. We have chosen $\left| \stackrel{\rightarrow }{e}%
_{i}\right| ^{2}=2$. The allowed momenta $\stackrel{\rightarrow }{p}$ take
values on the dual lattice of $\Lambda ^{D}$

\begin{equation}
\stackrel{\rightarrow }{p}=\sum_{i=1}^{D}m_{i}\left( \frac{1}{R_{i}}\sqrt{2}%
\stackrel{\rightarrow }{e}_{i}^{\star }\right) \in \left( \Lambda
^{D}\right) ^{\star }  \eqnum{3.3}
\end{equation}
The basis of $\left( \Lambda ^{D}\right) ^{\star }$ is $\left\{ \frac{1}{R1}%
\sqrt{2}\stackrel{\rightarrow }{e}_{1}^{\star },\frac{1}{R_{2}}\sqrt{2}%
\stackrel{\rightarrow }{e}_{2}^{\star },\ldots ,\frac{1}{R_{D}}\sqrt{2}%
\stackrel{\rightarrow }{e}_{D}^{\star }\right\} $ and we have $\stackrel{%
\rightarrow }{e}_{i}\cdot \stackrel{\rightarrow }{e}_{i}^{\star }=\delta
_{ij}$. The mode expansion of the compatified coordinates is

\begin{eqnarray}
\stackrel{\rightarrow }{X}_{R} &=&\frac{1}{2}\stackrel{\rightarrow }{x}%
+\left( \stackrel{\rightarrow }{p}-\frac{1}{2}\stackrel{\rightarrow }{L}%
\right) \left( \tau -\sigma \right) +i\sum_{r\neq 0}\frac{1}{r}\alpha
_{r}^{25}e^{-ir\left( \tau -\sigma \right) }  \eqnum{3.4} \\
\stackrel{\rightarrow }{X}_{L} &=&\frac{1}{2}\stackrel{\rightarrow }{x}%
+\left( \stackrel{\rightarrow }{p}+\frac{1}{2}\stackrel{\rightarrow }{L}%
\right) \left( \tau +\sigma \right) +i\sum_{r\neq 0}\frac{1}{r}\alpha
_{r}^{25}e^{-ir\left( \tau +\sigma \right) }  \eqnum{3.5}
\end{eqnarray}
The right and left momenta are defined to be $\stackrel{\rightarrow }{p}%
_{R}=\left( \stackrel{\rightarrow }{p}-\frac{1}{2}\stackrel{\rightarrow }{L}%
\right) $ and $\stackrel{\rightarrow }{p}_{L}=\left( \stackrel{\rightarrow }{%
p}+\frac{1}{2}\stackrel{\rightarrow }{L}\right) $. It can be shown that the
2D-vector $\left( \stackrel{\rightarrow }{p}_{R},\stackrel{\rightarrow }{p}%
_{L}\right) $ build an even self-dual Lorentzian lattice $\Gamma _{D,D}$,
which guarantees the string one loop modular invariance of the theory \cite
{8}. The moduli space of the theory is \cite{2}

\begin{equation}
\mu =\frac{SO\left( D,D\right) }{SO\left( D\right) \times SO\left( D\right) }%
/O\left( D,D,Z\right)  \eqnum{3.6}
\end{equation}
where $O(D,D,Z)$ is the discrete T-duality group and $dim$ $\mu =D^{2}$. To
complete the parametrization of the moduli space, one needs to introduce an
antisymmetric tensor field $B_{ij}$ in the bosonic string action. This will
modify the right (left) momenta to be

\begin{equation}
\stackrel{\rightarrow }{p}_{R}=\left( \stackrel{\rightarrow }{p}_{B}-\frac{1%
}{2}\stackrel{\rightarrow }{L}\right)  \eqnum{3.7}
\end{equation}

\begin{equation}
\stackrel{\rightarrow }{p}_{L}=\left( \stackrel{\rightarrow }{p}_{B}+\frac{1%
}{2}\stackrel{\rightarrow }{L}\right)  \eqnum{3.8}
\end{equation}
where

\begin{equation}
\stackrel{\rightarrow }{p}_{B}=\sum_{i,j}\left( m_{i}\frac{1}{R_{i}}\sqrt{2}%
\stackrel{\rightarrow }{e}_{i}^{\star }-n_{j}\frac{1}{\sqrt{2}R_{i}}B_{ij}%
\stackrel{\rightarrow }{e}_{i}^{\star }\right)  \eqnum{3.9}
\end{equation}
We are now ready to discuss the gauge state. As a first step, we restrict
ourselves to moduli space with $B_{ij}=0$ . For the type $I.a$ state, the $%
m=0$ constraint of Eq. (2.10) for massless states gives

\begin{equation}
N+\stackrel{\thicksim }{N}+\stackrel{\rightarrow }{p}^{2}+\frac{1}{4}%
\stackrel{\rightarrow }{L}^{2}=1  \eqnum{3.10}
\end{equation}

\begin{equation}
N-\stackrel{\thicksim }{N}=\sum_{i}m_{i}n_{i}-1  \eqnum{3.11}
\end{equation}
It is easy to see $N+\stackrel{\thicksim }{N}=0$ or $1$. For $N+\stackrel{%
\thicksim }{N}=1$, $m_{i}=n_{i}=0$, we have trivial gauge state solutions.
SGS exists for the case $N+\stackrel{\thicksim }{N}=0$ and the following
moduli points

\begin{equation}
R_{i}=\sqrt{2},e_{i}^{I}=\sqrt{2}\delta _{i}^{I}\text{ }\left( i=1,2,\ldots
,d\right)  \eqnum{3.12}
\end{equation}
with $m_{i}=n_{i}=\pm 1$, and $m_{j}=n_{j}=0$ for $d<j\leq D$. In each case,
the gauge states and SGS form a representation of $SU(2)^{d}$ algebra.
Similar results can be easily obtained for the type $I.b$ SGS. As in section
II, there is no massless type II SGS. We now discuss $B_{ij}\neq 0$ case.
For illustration, we choose $D=2$. In this case $B_{ij}=B\epsilon _{ij}$,
and one has four moduli parameters $R_{1},R_{2},B,$ and $\stackrel{%
\rightarrow }{e}_{1}\cdot \stackrel{\rightarrow }{e}_{2}$. For type $I.a$
state, the $m=0$ constraint of Eq. (2.10) gives

\begin{equation}
N+\stackrel{\thicksim }{N}+\stackrel{\rightarrow }{p}_{B}^{2}+\frac{1}{4}%
\stackrel{\rightarrow }{L}^{2}=1  \eqnum{3.13}
\end{equation}

\begin{equation}
N-\stackrel{\thicksim }{N}=m_{1}n_{1}+m_{2}n_{2}-1  \eqnum{3.14}
\end{equation}
SGS exists only for $N+\stackrel{\thicksim }{N}=0$. For the moduli point

\begin{equation}
R_{1}=R_{2}=\sqrt{2},B=\frac{1}{2},\stackrel{\rightarrow }{e}_{1}=\left( 
\sqrt{2},0\right) ,\stackrel{\rightarrow }{e}_{2}=\left( -\sqrt{\frac{1}{2}},%
\sqrt{\frac{3}{2}}\right)  \eqnum{3.15}
\end{equation}
one gets six SGS with momenta $\stackrel{\rightarrow }{p}_{R}$ being the six
root vectors of $SU(3)_{R}$. Together with two other trivial gauge states
corresponding to $N=0,\stackrel{\thicksim }{N}=1$ , they form the
Frenkel-Kac-Segal \cite{9} representation of $SU(3)_{k=1}$ Kac-Moody
algebra. Note that $\stackrel{\rightarrow }{e}_{1},\stackrel{\rightarrow }{e}%
_{2}$ are the two simple roots of $SU(3)$ and $\stackrel{\rightarrow }{e}%
_{1}^{\star }=\left( \sqrt{\frac{1}{2}},\sqrt{\frac{1}{6}}\right) ,\stackrel{%
\rightarrow }{e}_{2}^{\star }=\left( 0,\sqrt{\frac{2}{3}}\right) $. The six
sets of winding number are $\left( m_{1},n_{1},m_{2},n_{2}\right) =\left(
1,1,0,0\right) ,\left( -1,-1,0,0\right) ,\left( 0,0,1,1\right) ,\left(
0,0,-1,-1\right) ,\left( 1,1,1,0\right) ,\left( -1,-1,-1,0\right) $. Similar
results can be obtained for type $I.b$ SGS. The gauge states (including SGS)
thus form a representation of enhenced $SU(3)_{R}\otimes $ $SU(3)_{L}$ at
the moduli point of Eq. (3.15). In general, we expect that all enhenced
Kac-Moody gauge symmetry at any moduli point should have a realization on
SGS.

\section{ Massive Soliton Gauge State}

In this section we derive the massive SGS at the first massive level $M^{2}=2
$. We will find that SGS exists at infinite number of moduli points. One can
also show that they exist at an infinite number of massive level. The
existence of these massive SGS implies that there is an infinite enhanced
gauge symmetry structure of compatified string theory. For type $I.a$ state,
the $m=0$ constraint of Eq. (2.10) gives

\begin{equation}
\frac{m^{2}}{R^{2}}+\frac{1}{4}n^{2}R^{2}+N+\stackrel{\thicksim }{N}=3 
\eqnum{4.1}
\end{equation}

\begin{equation}
N-\stackrel{\thicksim }{N}=mn-1  \eqnum{4.2}
\end{equation}
which implies $N+\stackrel{\thicksim }{N}=0,1,2,3$. Equations (4. 1) and
(4.2) can be easily solved as following:

1. $N+\stackrel{\thicksim }{N}=3:$

\begin{equation}
m=n=0,N=1,\stackrel{\thicksim }{N}=2,\text{{\it any R}}  \eqnum{4.3}
\end{equation}

2. $N+\stackrel{\thicksim }{N}=2:$

\begin{eqnarray}
mn &=&1,N=\stackrel{\thicksim }{N}=1,R=\sqrt{2}  \nonumber \\
mn &=&-1,N=0,\stackrel{\thicksim }{N}=2,R=\sqrt{2}  \eqnum{4.4}
\end{eqnarray}

3. $N+\stackrel{\thicksim }{N}=1:$

\begin{eqnarray}
mn &=&2,N=1,\stackrel{\thicksim }{N}=0,R=2,1.\text{ }\left( T-duality\right)
\nonumber \\
mn &=&0,N=0,\stackrel{\thicksim }{N}=1,R=\frac{\left| m\right| }{\sqrt{2}},%
\frac{2\sqrt{2}}{\left| m\right| }.\text{ }\left( T-duality\right) 
\eqnum{4.5}
\end{eqnarray}

4. $N+\stackrel{\thicksim }{N}=0:$

\begin{equation}
mn=1,N=\stackrel{\thicksim }{N}=1,R=2\pm \sqrt{2}.\text{ }\left(
T-duality\right)  \eqnum{4.6}
\end{equation}
where we have included a T-duality transformation $R\rightarrow \frac{2}{R}$
for some moduli points. Note that Eq. (4.5) tells us that massive SGS exists
at an infinite number of moduli point. For type $II.a$ state, the $m=0$
constraint of Eq. (2.11) gives

\begin{equation}
\frac{m^{2}}{R^{2}}+\frac{1}{4}n^{2}R^{2}+N+\stackrel{\thicksim }{N}=2 
\eqnum{4.7}
\end{equation}

\begin{equation}
N-\stackrel{\thicksim }{N}=mn-2  \eqnum{4.8}
\end{equation}
which implies $N+\stackrel{\thicksim }{N}=0,1,2$. Equations (4.7) and (4.8)
can be solved as following:

1. $N+\stackrel{\thicksim }{N}=2:$

\begin{equation}
m=n=0,N=0,\stackrel{\thicksim }{N}=2,\text{{\it any R}}  \eqnum{4.9}
\end{equation}

2. $N+\stackrel{\thicksim }{N}=1:$

\begin{equation}
mn=1,N=0,\stackrel{\thicksim }{N}=1,R=\sqrt{2}  \eqnum{4.10}
\end{equation}

3. $N+\stackrel{\thicksim }{N}=0:$

\begin{equation}
mn=2,N=\stackrel{\thicksim }{N}=0,R=2,1.\text{ }\left( T-duality\right) 
\eqnum{4.11}
\end{equation}
The vertex operators of all SGS can be easily calculated and written down.
Similar results can be obtained for type $b$ gauge state. One can also
calculate propagating soliton states by using the same technique. We
summarize the moduli points which exist soliton state and SGS as following:

\mathstrut a. {\it Soliton gauge state :}

\begin{equation}
R=\sqrt{2},2\pm \sqrt{2},\frac{\left| m\right| }{\sqrt{2}},\frac{2\sqrt{2}}{%
\left| m\right| },2,1  \eqnum{4.12}
\end{equation}

\mathstrut b. {\it Soliton state :}

\begin{equation}
R=\sqrt{2},2\pm \sqrt{2},\frac{\left| m\right| }{\sqrt{2}},\frac{2\sqrt{2}}{%
\left| m\right| },\frac{\left| m\right| }{2},\frac{4}{\left| m\right| } 
\eqnum{4.13}
\end{equation}
In Eqs. (4.12) and (4.13), $m\in Z_{+}$. There is one interesting remark we
would like to point out by the end of this section. One notes that in the
second case of Eq. (4.5), instead of specifying $M^{2}=2$, in general we have

\begin{equation}
\frac{m^{2}}{R^{2}}+\frac{1}{4}n^{2}R^{2}=M^{2}  \eqnum{4.14}
\end{equation}
with $mn=0$. For say $R=\sqrt{2}$, one gets $M^{2}=\frac{m^{2}}{2}\left(
n=0\right) $. This means that we have an infinite number of massive SGS at
any higher massive level of the spectrum. One can even explicitly write down
the vertex operators of these SGS. We conjecture that the $w_{\infty }$
symmetry of 2D string theory \cite{6,10} can be realized in these SGS \cite
{11}. Other moduli points also consist of higher massive SGS in the spectrum.

\section{Conclusion}

It is hoped that all space-time symmetry of string theory are due to the
existence of gauge state in the spectrum. The Heterotic gauge state for the
10D Heterotic string and discrete gauge state for the toy 2D string are such
examples. We have introduced soliton gauge state (SGS) for compatified
string in this paper, and have related them to the enhanced Kaluza-Klein
Kac-Moody gauge symmetry in the theory. In many cases, especially for the
massive states, it is easier to study gauge symmetry in the gauge state
sector than in the propagating spectrum directly. Since the discrete
T-duality symmetry group for bosonic string is the Weyl subgroup of the
enhanced gauge group, it can also be considered as implied by the existence
of SGS. It is not clear whether other discrete duality symmetry group can be
understood in this way. Finally, it would be interesting to consider more
complicated compatification, e.g. orbifold and Calabi-Yau compatifications
and study the relation between SGS and duality symmetry. works in this
direction is in progress.

\section{Acknowledgments}

This research is supported by National Science Council of Taiwan, R.O.C.,
under grant number NSC86-2112-M009-016.

\end{document}